\documentclass{lmcs}

\usepackage{amsmath}
\usepackage{amssymb}
\usepackage{graphicx}
\usepackage{bm}
\usepackage{bbold}
\usepackage{babel}
\usepackage[utf8]{inputenc}
\usepackage[T1]{fontenc}

\title[Probabilistic Shoenfield Machines]{Probabilistic Shoenfield Machines}

\author[M.~Bujok]{Maksymilian Bujok\rsuper a}
\address{\lsuper a Faculty of Design, SWPS University, Chodakowska 19/31, 03-815 Warsaw, Poland}
\email{mbujok@swps.edu.pl}

\author[A.~Mata]{Adam Mata\rsuper b}
\address{\lsuper b Faculty of Mathematics and Information Science, Warsaw University of Technology, Koszykowa 75, 00-662 Warsaw, Poland}
\email{adam.mata.dokt@pw.edu.pl}


\begin{document}

\begin{abstract}
The article provides the theoretical framework of Probabilistic Shoenfield Machines (\textbf{PSMs}), an extension of the classical Shoenfield Machine that models randomness in the computation process. PSMs are introduced in contexts where deterministic computation is insufficient, such as randomized algorithms. By allowing transitions to multiple possible states with certain probabilities, PSMs can solve problems and make decisions based on probabilistic outcomes, thus expanding the variety of possible computations. We provide an overview of PSMs, detailing their formal definitions, the computation mechanism, and their equivalence with Non-deterministic Shoenfield Machines (NSMs).
\end{abstract}

\maketitle

\section{Abstract}

The  article provides the theoretical framework of Probabilistic Shoenfield Machines (\textbf{PSMs}), an extension of the classical Shoenfield Machine that models randomness in the computation process. \textbf{PSMs} are brought in contexts where deterministic computation is insufficient, such as randomized algorithms. By allowing transitions to multiple possible states with certain probabilities, \textbf{PSMs} can solve problems and make decisions based on probabilistic outcomes, hence expanding the variety of possible computations. We provide an overview of \textbf{PSMs}, detailing their formal definitions as well as the computation mechanism and their equivalence with Non-deterministic Shoenfield Machines (\textbf{NSM}).

\section{Introduction}

Theoretical computer science is rich with different models that were created to formalize and understand the limits of computation. Among these, the Deterministic Shoenfield Machine (\textbf{DSM}) turns out to be the simplest one, providing a fundamental framework to recognize whether a function is computable or not. However, as the work in that matter progressed, the necessity occurred to extend this classical model to capture a broader spectrum of computational phenomena, creating the probabilistic models of computation. One of these significant extensions, which is provided in this paper, is the Probabilistic Shoenfield Machine (\textbf{PSM}). In this article, we discuss what PSM is, stating its formal definition and mechanism of computation.

A Probabilistic Shoenfield Machine is a type of the \textbf{DSM} that models the elements of randomness in its computation process. This model is particularly pertinent in contexts where deterministic computation may not be applicable, such as in the case of randomized algorithms, cryptographic protocols, and probabilistic analysis. Unlike a classical Shoenfield Machine, which computes along a single path defined by its deterministic transition function, a \textbf{PSM} can transition into multiple possible states with different probabilities. This probabilistic behavior allows PSMs to solve problems and make decisions based on different outcomes, thus adding to the spectrum of things that can be computed.

The concept of \textbf{PSMs} not only broadens the number of theoretical means of computation but also has practical implications. For example, probabilistic algorithms, which are exploited by many modern technologies, rely on the probabilistic computation schema to achieve faster or more efficient solutions compared to their deterministic computations. Moreover, the study of such models interplays with other branches of theoretical computer science.

By extending the computational potential of traditional Shoenfield Machines to include probabilistic transitions, \textbf{PSMs} represent an advancement in the theoretical understanding of computation. Through this study, we hope to present aspects of probabilistic computation and its impact on the development of the study of computational matter.

The following document provides introductions of Non-deterministic Shoenfield Machines (\textbf{NSM}) and Probabilistic Shoenfield Machines (\textbf{PSM}), both based on Deterministic Shoenfield Machines {\textbf{DSN}, as a formal model of computation. The document begins with a detailed definition of DSM, outlining its components and the Representation of its computation. It further shows a short introduction of \textbf{NSM} and establishes its computational equivalence with \textbf{DSM}. Next, the \textbf{PSM} is introduced, and its equivalence to DSM is discussed.

\section{Deterministic Shoenfield Machines}

A Deterministic Shoenfield Machine (\textbf{DSM}) is a formal model of computation introduced by J.~R.~Shoenfield in~\cite{Shoenfield} and then further simplified by Y.~L.~Yershov in~\cite{Yershov}. In this paper, we exploit the fact that it was shown that the computational power of the model is equivalent to that of Turing Machines.

\subsection{Definition of \textbf{DSM}}

A Deterministic Shoenfield Machine is a computation model composed of:

\begin{itemize}
    \item An infinite set of \textbf{registers} enumerated by natural numbers $0,1,2,\ldots$. Each register $R_i$ is a memory cell containing a natural number. The number in a particular register may change its value during the computation. During the computation, the machine uses only a finite number of registers. The purpose of registers is to store the data (natural numbers).
    \item An \textbf{instruction counter}. It is a memory cell containing a natural number at any time. The number points to the index of the instruction within the program which is to be executed next. At the beginning of the computation, it contains $0$.
    \item A program placed in a separate part of the machine's memory. It is a \textbf{finite} list of instructions enumerated from $0$ to some $n$. During the computation, the program \textbf{does not} change.
\end{itemize}

The program is written before running the \textbf{DSM}, registers are filled with input data, and the instruction counter is set to $0$. The machine executes the instruction at one step, the index of which is present in the instruction counter. The machine stops its computation only if the instruction counter points to an instruction number that does not exist in the program. It is also possible that a machine never halts.

There are only two types of instructions:
\begin{itemize}
    \item \textbf{$\mathsf{INC}\ i$} -- during the execution of this instruction, the machine increments the value stored in the $i$-th register and the instruction counter by one. Then, the machine continues to the next step.
    \item \textbf{$\mathsf{DEC}\ i,n$} -- if at the beginning of the execution of this instruction, the value stored in the $i$-th register is greater than $0$, then the machine decrements this value by one and sets the instruction counter to $n$. Otherwise, if the value stored in the $i$-th register equals $0$, then the machine increments the instruction counter by one.
\end{itemize}

\subsection{Representation of Computation of a DSM}

A temporary configuration of a particular \textbf{DSM} may be represented by two pieces of data:

\begin{itemize}
    \item current values stored in all of the \textbf{registers},
    \item current value of the \textbf{instruction counter}.
\end{itemize}

Let us note that we may represent a computation of a particular \textbf{DSM} as a descending chain (possibly infinite) where the top element represents the initial configuration. The direct descendant of the top element is the configuration obtained after executing the program instruction which is present in line number $0$ (since, at the beginning of the computation, the instruction counter value is $0$). The following configurations in the chain are constructed as resulting configurations of executing the instruction pointed to by an instruction counter on the preceding configuration.

\begin{figure}[htbp]
    \centering
    \includegraphics[scale=0.9]{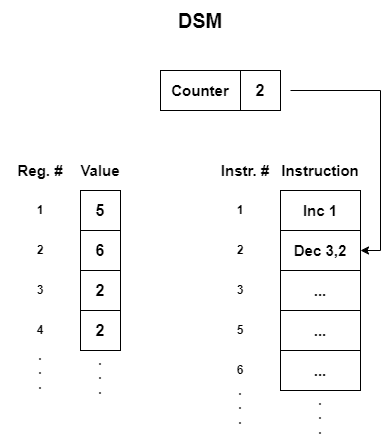}
    \caption{This figure represents a Deterministic Shoenfield Machine (\textbf{DSM}). The machine has a counter set to $2$, which points to the next instruction. Multiple registers hold integer values: Register 1 has $5$, Register 2 has $6$, Register 3 has $2$, and Register 4 has $2$. The instruction list includes operations such as $\mathsf{INC}\ 1$ (increment Register 1) and $\mathsf{DEC}\ 3,2$ (decrement Register 3 and set the counter to $2$ if the result is zero). The current instruction to be executed is $\mathsf{DEC}\ 3,2$.}
    \label{fig:DSM}
\end{figure}

\section{Non-deterministic Shoenfield Machines}

In this section, we introduce Non-deterministic Shoenfield Machines (\textbf{NSM}), and we show their computational equivalence with \textbf{DSM}. 

\subsection{Definition of \textbf{NSM}}
A Non-deterministic Shoenfield Machine (\textbf{NSM}) is a computation model composed of:

\begin{itemize}
    \item An infinite array of registers, as in the case of \textbf{DSM}.
    \item An instruction counter. The instruction counter is set to $0$ at the beginning of the computation.
    \item A program placed in a separate memory part of the machine as in the case of \textbf{DSM}; however, each program line contains a \textbf{finite number} of instructions.
\end{itemize}

The program is written before running the \textbf{NSM}, registers are filled with the input data, and the instruction counter is set to $0$. At one step, the machine executes the program line, which is pointed to by the instruction counter. If there is more than one instruction in the cell, the machine replicates the current configuration of the registers, counter and the program, creating as many copies of the machine as the number of instructions in the current program line. Then, the machine executes one instruction from the line separately in each copy and updates the instruction counters according to the instruction executed. The computation in each cell proceeds separately. The machine halts if all of the submachines created halt. Otherwise, it works without halting. 

\begin{figure}[htbp]
   \centering
   \includegraphics[scale=0.8]{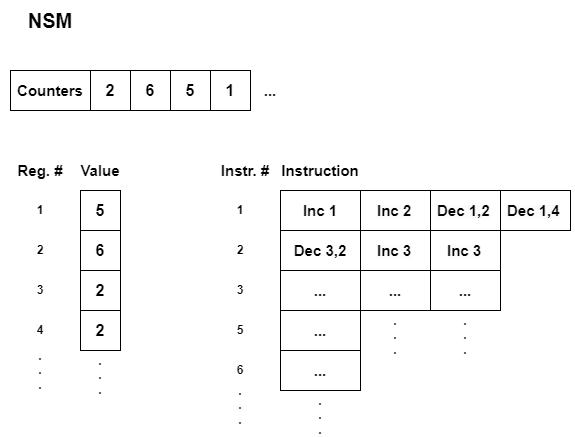}
   \caption{This diagram represents a Non-deterministic Shoenfield Machine (\textbf{NSM}). The machine consists of multiple counters ($2,6,5,1,\ldots$), which point to possible next instructions. Registers hold integer values: Register~1 has~$5$, Register~2 has~$6$, Register~3 has~$2$, and Register~4 has~$2$. The instruction list includes operations such as $\mathsf{INC}\ 1$, $\mathsf{INC}\ 2$, $\mathsf{DEC}\ 1,2$, $\mathsf{DEC}\ 1,4$, and $\mathsf{DEC}\ 3,2$. This machine can follow different execution paths due to multiple counters pointing to different instructions.}
   \label{fig:NSM}
\end{figure}

\subsubsection{Theorem}
Every \textbf{NSM} is equivalent to some \textbf{DSM}.

\textit{Proof.}
Let $M$ be an \textbf{NSM}. First, we present a procedure for building a \textbf{tree $T$ of the computation of $M$}. Second, we show that every path from the root of $T$ to any of its leaves represents a finite computation of a \textbf{DSM}. If some path is infinite, it represents an infinite computation of some \textbf{DSM}.

\subsection{Building up the computation tree $T$ of NSM $M$}

\textbf{1)} Let $k_0$ be an initial configuration of registers of machine $M$. Let $n_0$ be an initial instruction counter value of machine $M$. In the root of $T$ we store information about $k_0$ and $n_0$:

$$
k_0 \mid n_0 := w_0
$$

\textbf{2)} Now, let us assume that our starting node is:
$$
w_i = k_i \mid n_i
$$
The succeeding nodes of $T$ are constructed inductively as follows: 

\textbf{A)} If $n_i$ points at the line of program which contains \textbf{only one} instruction $c$ then the only descendant of the node $w_i$ is the node $w_{i+1} = k_{i+1} \mid n_{i+1}$, where $k_{i+1}$ is the configuration of registers obtained after execution of $c$ on configuration $k_i$. Further, $n_{i+1}$ is the value of the instruction counter we obtain after executing $c$, assuming that the previous value of the instruction counter was $n_i$.

\textbf{B)} If $n_i$ points at the line of program which contains several instructions $c_1,c_2,\ldots,c_m$ then there are created $m$ descendants of the node $w_i$:

$$
w^1_{i+1} = k^1_{i+1} \mid n^1_{i+1},\ w^2_{i+1} = k^2_{i+1} \mid n^2_{i+1},\ \ldots,\ w^m_{i+1} = k^m_{i+1} \mid n^m_{i+1}
$$

where $w^a_{i+1}$ is obtained by executing $c_a$ on the configuration $w_i$ in the same manner as in \textbf{A)}.

According to the rules of building the tree $T$, there are two options for how each of the branches of the tree may look:

\begin{itemize}
    \item The path may end with a final configuration, which we call a \textbf{leaf node}.
    \item The path may be infinite.
\end{itemize}

Let us notice that every path from the root of $T$, which ends in a leaf $w_n$, is equivalent to the computation of some \textbf{DSM}. Based on transitions between each node and its direct descendant in the particular path, we can recover instructions executed along the path. That allows us to recover the program.

In the case of the infinite path, we can also recover a program executed along the path. As long as the initial program of $M$ is finite, the one executed in the path must be finite and looped so it executes infinitely.

If the path in $T$ is finite then, by \cite[Theorem~34]{AlgorithmsKogabaev} it is equivalent to a computation of some partial recursive function and hence, such a path is also equivalent to some \textbf{DSM} for which the input $w_0$ results in a defined value. 

If the path in $T$ is infinite, then, by the same theorem from \cite{AlgorithmsKogabaev}, it is equivalent to a computation of some \textbf{DSM} for which the input $w_0$ results in infinite computation.

\begin{figure}[htbp]
   \centering
   \includegraphics[scale=0.8]{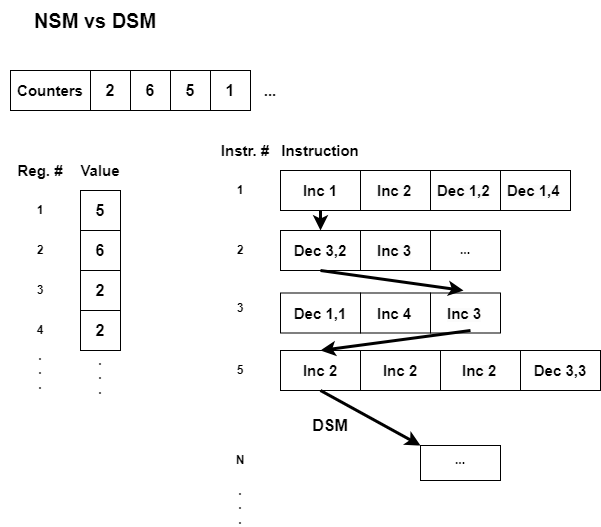}
   \caption{This diagram compares a Non-deterministic Shoenfield Machine (\textbf{NSM}) and a Deterministic Shoenfield Machine (\textbf{DSM}). The \textbf{NSM} has multiple counters ($2,6,5,1,\ldots$) pointing to possible next instructions, with registers holding values $5$, $6$, $2$, and $2$. Instructions include operations such as $\mathsf{INC}\ 1$, $\mathsf{DEC}\ 3,2$, and $\mathsf{DEC}\ 1,2$, with arrows showing possible transitions. The \textbf{DSM} follows a single deterministic path between instructions.}
   \label{fig:NSMvsDSM}
\end{figure}

This completes the proof of equivalence between \textbf{NSM} and \textbf{DSM}.
\section{Probabilistic Shoenfield Machines}
In this section, we introduce Probabilistic Shoenfield Machines (\textbf{PSM}), and we show their computational equivalence with \textbf{DSM} (\text{NSM}, TM). By design, the probabilistic Shoenfield machine is a development of the non-deterministic Shoenfield machine, where the choice of transition between specific states is made on the basis of a particular probability distribution.

\subsection{Definition of \textbf{PSM}}

\begin{itemize}
    \item An infinite array of cells containing an infinite set of registers, as in the case of \textbf{DSM} and \textbf{NSM}.
    \item An infinite array of instruction counters. The first instruction counter is set to 0 at the beginning of the computation, while the value of the others is set to blank.
    \item A program is placed in a separate memory part of the machine as in the case of \textbf{DSM} and \textbf{NSM}. Each program line contains a \textbf{finite number} of instructions.
    \item A non-uniform oracle is a mechanism that generates random values from a specified set, where each value has an assigned probability. In the context of the defined machine with $n$ instructions in each row, selects an instruction based on a given set of probabilities $ \{ p_1, p_2, \ldots, p_n \} $, where $\sum_{i=1}^n p_i = 1$  and $i$ is the number of instructions in the line.
    \item  The execution of a given instruction in a line is done with the selected probability $p_i$. 
\end{itemize}

What else can we write down in the form of:

\textbf{Definition (PSM).}\\
  A probabilistic machine is a quadruple:
  \[
    M = (\mathcal C,\;c_0,\;F,\;\delta)
  \]
  where:
  \begin{itemize}
    \item \( \mathcal C \) – the set of configurations (register values + program counter),
    \item \( c_0 \in \mathcal C \) – the initial configuration,
    \item \( F \subseteq \mathcal C \) – the set of accepting configurations,
    \item \( \delta:\mathcal C \longrightarrow \textsf{Dist}(\mathcal C) \) – the probabilistic transition function, such that \(\sum_{c'} \delta(c, c') = 1\).
  \end{itemize}

  \textbf{Theorem 1 (Decidability).}\\
  
  For every PSM \( M \) and every \( \varepsilon \in (0,1) \), there exists a deterministic machine (DSM) that, for any input \( x \), computes a value \( q \) such that:
  \[
    \bigl|\,\Pr_M(x) - q\,\bigr| \;\le\; \varepsilon.
  \]

  \vspace{0.5em}
  \textit{Proof – outline.}
  \begin{enumerate}
    \item Let \( r(n) \) be the maximal number of coin tosses performed by \( M \) during the first \( T(n) \) steps (i.e., the runtime of PSM on inputs of length \( n \)).
    \item The DSM enumerates all bitstrings \( w \in \{0,1\}^{r(n)} \) (there are \( 2^{r(n)} \) such strings), and deterministically simulates \( M \) for each \( w \).
    \item Counting the fraction of accepting runs gives the exact value of \( \Pr_M(x) \).\\
    Alternatively, sampling only the first \( k = \lceil (1/\varepsilon^2) \ln 3 \rceil \) random bitstrings and applying the Chebyshev/Chernoff inequality yields an approximation within error \( \le \varepsilon \).
  \end{enumerate}

  \vspace{0.5em}
  \textit{Complexity.} The runtime of the DSM is \( O(2^{r(n)}) \) for the exact version, and \( O((1/\varepsilon^2) \cdot r(n)) \) for the randomized variant with error control.
}

Analogous to NSM, before the PSM is run, the program is written, the registers in the first cell of the array are filled with input data, and the counter of the first instruction is set to 0. The machine executes a program line in one step indicated by the instruction counter. However, in the case of PSM, the execution of a given instruction in a line is done with the probability $p_i$. The sum of all right-likelihoods in a line, $\sum_{i=1}^{n}p_i=1$, where $i$ is the instruction number, and $n$ is the number of instruction in line. The $p_i$ probabilities are not necessarily equal. 

It is worth noting that the execution of each instruction with a certain probability is the probability of a given configuration of registers after the operation.

In contrast to the NSM, the machine can only execute one instruction per line with a probability of $p_i$.

The probability of moving to any state beyond the current one is zero if and only if it is an accepting state (final state).

\subsection{Equivalence between PSM and NSM}

The equivalence between NSM and PSM is due to the fact that the computation in PSM is the same as in NSM, and a particular instruction is selected with probability $p_i$, where the probability of selecting all instructions in a line adds up to 1. From this, the equivalence between PSM and NSM is shown.

\subsection{Bulding up the computation path P of PSM}

Unlike NSM, in the case of PSM, we are dealing with a path P in the tree.

\textbf{1)} Let $k_0$ be an initial configuration of registers of machine $M$. Let $n_0$ be an initial instruction counter value of machine $M$. In the 
first row of $P$ we store information about $k_0$ and $n_0$:

$$
k_0 | n_0 := w_0.
$$

\noindent\textbf{2)} Now, let us assume that our starting node or path is:
$$
w_i = k_i | n_i.
$$
The succeeding nodes of $P$ are constructed inductively as follows:

\textbf{A)} If $n_i$ points at the line of the program which contains \textbf{only one} instruction $c$, whose execution probability is 1, then, as with NSM, the only descendant of the node $w_i$ is the node $w_{i + 1} = k_{i + 1} | n_{i + 1}$, where $k_{i + 1}$ is the configuration of registers obtained after execution of $c$ on configuration $k_i$. Further, $n_{i + 1}$ is the value of the instruction counter we obtain after executing $c$, assuming that the previous value of the instruction counter was $n_i$.

\textbf{B)} If $n_i$ points to a program line that contains several instructions $c_1, c_2, \dotso, c_m$, then one of them is executed with probability $p_i$. In this case, however, unlike NSM, an execution path is created. Thus, one descendant of the node $w_i$ is created:

$$
w^a_{i + 1} = k^a_{i + 1} | n^a_{i + 1},
$$

where $w^a_{i + 1}$ is obtained by executing $c_a$ on the configuration $w_i$ in the same manner as in \textbf{A)}.

Thus, with probability $p_i$, one descendant of node $w_i$ is created. According to the presented $P$ path construction rules, there are two options for the appearance of each branch of the tree:

\begin{itemize}
    \item A path can end with a final configuration, called \textbf{leaf node}.
    \item The path can be infinite.
\end{itemize}

By analogy with \textbf{NSM}, let us note that each path from the root $P$ that ends in the leaf $w_n$ is equivalent to computing some \textbf{DSM}. We can recover the instructions executed along that path based on the transitions between each node and its direct descendant in a given path. 

The probability of executing an instruction $w^a_{i+1}$, which is one of the descendants of node $w_i$ in the Probabilistic Shoenfield Machine (PSM) model, depends on the probability of selecting a particular instruction $c_a$ at a given program stage. 

The single path of computation in PSM is equivalent to DSM, and therefore, the same principles apply as we discussed in the section on DSM. The only difference is that it is executed with a probability equal to the product of the probabilities of the drawn instructions.

\begin{center}
    \begin{figure}
       \includegraphics[width=\textwidth]{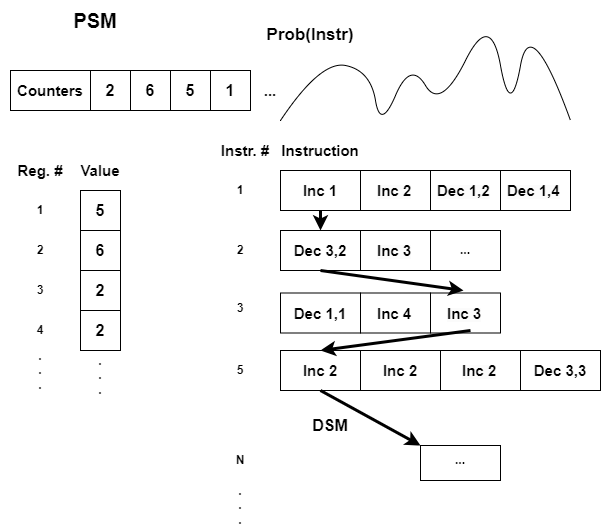}
       \caption{The Probabilistic Shoenfield Machine (PSM). In this case, the machine has multiple counters (2, 6, 5, 1, ...) indicating possible next instructions, with registers holding values 5, 6, 2, and 2. Instructions include operations such as Inc 1, Dec 3,2, and Dec 1,2, with arrows showing possible transitions. A graph represents the probabilities of different instructions. The DSM follows a single deterministic path between instructions.}
       \end{figure}
        \label{fig:PSM}
\end{center}

\subsection{Computational Equivalence with DSM and NSM}
\label{sec:psm_equivalence} 

While the previous sections defined the operational mechanics of \textbf{DSM}, \textbf{NSM}, and \textbf{PSM}, a crucial aspect is their relative computational power. We argue that, under appropriate acceptance definitions, these models are computationally equivalent. This means they can decide the same class of languages and compute the same class of functions, namely the partial recursive functions, establishing their equivalence to Turing Machines.

The equivalence between Deterministic and Non-deterministic Shoenfield Machines ($\textbf{NSM} \equiv \textbf{DSM}$) is a standard result in computability theory, demonstrated earlier by simulating the computation tree of an \textbf{NSM} with a \textbf{DSM}. We now extend this equivalence to include \textbf{PSM}.

\subsubsection{Simulating DSM and NSM using PSM}

Simulating a \textbf{DSM} using a \textbf{PSM} is straightforward. A deterministic computation is merely a special case of a probabilistic one where all transition probabilities are either 0 or 1. For any configuration $c$ where the \textbf{DSM} deterministically transitions to a unique next configuration $c'$, the corresponding \textbf{PSM} can be defined with a probabilistic transition function $\delta$ such that $\delta(c, c') = 1$ and $\delta(c, c'') = 0$ for all other configurations $c'' \neq c'$. This \textbf{PSM} will perfectly replicate the \textbf{DSM}'s computation path with probability 1.

Given that $\textbf{NSM} \equiv \textbf{DSM}$, the ability of a \textbf{PSM} to simulate a \textbf{DSM} directly implies that a \textbf{PSM} can also simulate the computational power of an \textbf{NSM}. Therefore, in terms of computability, the power of \textbf{PSM} is at least that of \textbf{DSM} and \textbf{NSM}:
\[
\textbf{DSM} \le \textbf{PSM} \quad \text{and} \quad \textbf{NSM} \le \textbf{PSM}.
\]

\subsubsection{Simulating PSM using DSM}

Demonstrating the reverse direction, that $\textbf{PSM} \le \textbf{DSM}$, requires careful consideration of how a \textbf{PSM} accepts or decides a language. We adopt the standard definition from complexity theory: \textbf{bounded-error probabilistic acceptance}. A language $L$ is decided by a \textbf{PSM} $M$ if $M$ halts with probability 1 on all inputs, and there exists a constant $\eta > 0$ (the error bound, defining a gap around $1/2$) such that for any input $x$:
\begin{itemize}
    \item If $x \in L$, then the probability that $M$ accepts $x$, denoted $\Pr(M \text{ accepts } x)$, satisfies $\Pr(M \text{ accepts } x) \ge \frac{1}{2} + \eta$. (Commonly, $\frac{1}{2} + \eta = \frac{2}{3}$).
    \item If $x \notin L$, then $\Pr(M \text{ accepts } x) \le \frac{1}{2} - \eta$. (Commonly, $\frac{1}{2} - \eta = \frac{1}{3}$).
\end{itemize}
The probability is calculated over the space of all possible sequences of random choices made by the machine's non-uniform oracle during the computation on $x$. We assume that the probabilities associated with the transitions in the \textbf{PSM} program are computable numbers (e.g., rational numbers), allowing a \textbf{DSM} to work with them.

Under this definition, a \textbf{DSM} can simulate the \textbf{PSM} $M$ to decide the same language $L$. This relies on the ability of the \textbf{DSM} to approximate the acceptance probability $p = \Pr(M \text{ accepts } x)$ with arbitrary accuracy $\varepsilon$, as outlined in Theorem~1 (Decidability). The \textbf{DSM} simulates a sufficiently large number $N$ of computation paths of the \textbf{PSM} $M$ on input $x$. Each path corresponds to a specific sequence of outcomes from the probabilistic choices. The \textbf{DSM} can generate these sequences deterministically (e.g., by iterating through all binary strings of a certain length representing the choices). By calculating the fraction $p_{approx}$ of these simulated paths that result in acceptance, the \textbf{DSM} obtains an approximation of $p$. Using concentration inequalities (like Chernoff bounds), it can be shown that for a sufficiently large $N$ (computable from $\eta$ and a desired confidence level), the approximation $p_{approx}$ will be close enough to $p$ such that $|p_{approx} - p| < \eta$.

The simulating \textbf{DSM} then decides as follows:
\begin{enumerate}
    \item Given input $x$, calculate the required number of simulations $N$ based on $\eta$.
    \item Perform $N$ deterministic simulations of $M$'s computation paths on $x$.
    \item Compute the approximate acceptance probability $p_{approx}$.
    \item If $p_{approx} > 1/2$, the \textbf{DSM} accepts $x$.
    \item Otherwise (if $p_{approx} \le 1/2$), the \textbf{DSM} rejects $x$.
\end{enumerate}
Because $|p_{approx} - p| < \eta$, if $x \in L$ (so $p \ge 1/2 + \eta$), then $p_{approx}$ must be greater than $1/2$. Conversely, if $x \notin L$ (so $p \le 1/2 - \eta$), then $p_{approx}$ must be less than or equal to $1/2$. This deterministic procedure correctly decides the language $L$. This demonstrates that any language decidable by a \textbf{PSM} with bounded error is also decidable by a \textbf{DSM}, establishing $\textbf{PSM} \le \textbf{DSM}$.

\subsubsection{Conclusion on Equivalence}

Combining the simulations in both directions ($\textbf{DSM} \le \textbf{PSM}$ and $\textbf{PSM} \le \textbf{DSM}$ under the bounded-error model), we conclude that \textbf{PSM} has the same fundamental computational power as \textbf{DSM}. Given the established equivalence $\textbf{NSM} \equiv \textbf{DSM}$, it follows that all three formalisms – Deterministic, Non-deterministic, and Probabilistic Shoenfield Machines – are computationally equivalent in terms of the class of languages they decide and the functions they compute. They all capture the class of partial recursive functions, equivalent to the power of Turing Machines.

\section{Aplications}

\subsection{Potential Applications for Probabilistic Computations}

Computer science applications cover areas where something more than classical computational models, such as the classical Turing Machine (TM), may be required. Modeling computation with a Probabilistic Turing Machine (PTM) may be more appropriate because the PTM considers elements of randomness, which is particularly important in environments where deterministic operating conditions cannot be guaranteed.

Such an environment is, for example, space and the equipment operating there, which must cope with intense radiation. Crossing Jupiter's radiation belts exposes hardware and algorithms to high radiation levels, which can disrupt traditional calculations. Under such conditions, the PTM can better reflect the natural operating environment, considering the effects of radiation on computational processes.

Under such conditions, the PTM may be better suited to study computing behavior in considerable measure by approximating the radiation-related conditions there.

A very similar situation exists in the case of nuclear remediation robots.

Even in everyday computations involving many individual operations, such as calculations of  $\pi$ with record precision, cosmic radiation can introduce unpredictable disturbances that influence the results. Using a Probabilistic Turing Machine (PTM) is invaluable in such cases. PTMs can account for these unpredictable disturbances, leading to more reliable outcomes. This versatility of PTMs in accounting for unpredictable factors in everyday computations is a crucial aspect of their application.
In summary, the Probabilistic Turing Machine and, per analogy, the Probabilistic Shoenfield Machine could find applications in many areas where more classical deterministic computational models, like TM, may not meet the challenges posed by unpredictable external factors. By incorporating elements of randomness, PTM allows for more realistic and reliable modeling of computational processes under challenging conditions. \cite{Draege}

\subsection{Potential Applications analog to PTM}


Meanwhile, PTMs, as a formalism, have well-established and well-known applications. However, by analogy and equivalence between PTMs and PSMs, we can expect them to perform well in many similar applications and areas, for example, Generating random numbers and keys \cite{Klingler} and Probabilistic encryption schemes \cite{Klingler}.

As another example, we can consider randomized algorithms, where PSM can provide a theoretical basis for randomized algorithms that use random choices to solve problems more efficiently than deterministic algorithms. In this case, examples include New methods for primality testing,  \cite{Arora}, and polynomial identity testing \cite{Arora}.

Particularly promising seems to be PSM's potential ability to model complex randomness in systems such as quantum systems, neural networks, or complex biological processes, for example.

\section{Conclusions}

\section{Conclusions and Future Work}

In this work, we have undertaken a systematic study of the Shoenfield machine, a foundational model of computation based on register machines. Building upon the classical Deterministic Shoenfield Machine (\textbf{DSM}), we introduced and formalized two significant extensions: the Non-deterministic Shoenfield Machine (\textbf{NSM}) and the Probabilistic Shoenfield Machine (\textbf{PSM}).

We provided rigorous definitions for both \textbf{NSM} and \textbf{PSM}. The \textbf{NSM} extends the \textbf{DSM} by allowing multiple instructions per program line, representing non-deterministic choices through the conceptual replication of machine states and parallel exploration of computation paths. The \textbf{PSM}, in turn, builds upon this structure by introducing a probabilistic element: instead of exploring all branches, a single instruction is chosen from a program line according to a specified probability distribution, modeling randomized computation within the Shoenfield framework.

A central theoretical contribution of this paper is the demonstration of the computational equivalence between these three models. We explicitly showed, through the construction and analysis of computation trees, that any computation performable by an \textbf{NSM} can be simulated by a \textbf{DSM}, confirming $\textbf{NSM} \equiv \textbf{DSM}$. Furthermore, we argued compellingly for the equivalence of the \textbf{PSM} with the \textbf{DSM} and \textbf{NSM}. This was established by showing that a \textbf{DSM} can simulate a \textbf{PSM} under the standard bounded-error acceptance criterion, primarily by approximating the acceptance probabilities (as formalized in Theorem~1), while a \textbf{PSM} can trivially simulate a \textbf{DSM} by using probabilities of only 0 and 1. This fundamental result reinforces the robustness of the class of computable functions (the partial recursive functions) and confirms that \textbf{DSM}, \textbf{NSM}, and \textbf{PSM} all share the computational power equivalent to that of Turing Machines.

The introduction of \textbf{NSM} and \textbf{PSM} offers alternative perspectives and potentially simpler frameworks, compared to Turing machine variants, for theoretical investigations into non-determinism and probabilistic computation. The \textbf{PSM}, in particular, provides a formal basis analogous to the Probabilistic Turing Machine (\textbf{PTM}) for analyzing randomized algorithms and modelling computational processes influenced by inherent randomness or environmental noise, such as those encountered in cryptography, complex system simulation, or computations in harsh environments like outer space. We have outlined several potential application areas, including random number generation, probabilistic encryption schemes, and algorithms for problems like primality testing.

This work opens several avenues for future research.
\begin{itemize}
    \item \textbf{Complexity Analysis:} While we focused on computability, a natural next step is to analyze the time and space complexity of computations on \textbf{NSM} and \textbf{PSM}. How do resource-bounded versions of these machines relate to standard complexity classes like P, NP, and BPP?
    \item \textbf{Specific Algorithms and Applications:} Developing and analyzing concrete algorithms for \textbf{PSM} for specific problems (e.g., optimization, machine learning tasks, identity testing) could yield valuable insights and comparisons with existing PTM-based or classical algorithms.
    \item \textbf{Model Variations:} Exploring variants of the \textbf{PSM} model could be fruitful. For instance, what is the impact of different classes of probability distributions accessible via the oracle? Could multi-tape or multi-counter versions offer advantages?
    \item \textbf{Quantum Shoenfield Machines:} Extending the Shoenfield framework further to incorporate quantum mechanics, perhaps defining a "Quantum Shoenfield Machine" analogous to Quantum Turing Machines or building upon existing work on Quantum Register Machines \cite{leporati2007}, presents an intriguing direction.
    \item \textbf{Implementation and Education:} Developing simulators for \textbf{DSM}, \textbf{NSM}, and \textbf{PSM} could serve as valuable educational tools for teaching fundamental concepts of computability, non-determinism, and probabilistic computation.
\end{itemize}

In summary, by defining and establishing the equivalence of the Non-deterministic and Probabilistic Shoenfield Machines, this paper contributes to the foundational theory of computation and provides versatile models for further theoretical exploration and potential application.


\end{document}